\documentclass[letterpaper]{jpconf}    
\usepackage{graphicx}    
\input babarsym 
\input hoobersym 
\usepackage{times}
\begin{document}
\title{Searches for Exotic Decays of the \ups\ at \lbabar}

\author{Benjamin Hooberman, on behalf of the \babar\ collaboration}

\address{Lawrence Berkeley  National Laboratory and  the University of
California at Berkeley}

\begin{abstract}
In this paper we present two  searches for new physics in \ups\ decays
collected  by  the  \babar\   detector.   We  search   for  charged
lepton-flavour violating decays of the \ups, which are unobservable in
the   Standard  Model   but  are   predicted  to   occur   in  several
beyond-the-Standard Model scenarios.  We also search for production of
a light Higgs or Higgs-like  state produced in radiative decays of the
\ups\ and decaying to muon pairs.
\end{abstract}

\section{Introduction}
When  the Large  Hadron Collider  becomes operational,  it  will probe
physics  beyond  the  Standard  Model  (SM) by  searching  for  direct
production of  new particles such as those  predicted by Supersymmetry
and models  with extra  dimensions.  However, it  is also  possible to
search for new  physics in data collected at  lower collision energies
by searching for rare and  exotic processes which are forbidden in the
SM.  In  this paper  two such searches  are presented, which  use data
collected  by the \babar\  detector~\cite{cite_babar} situated  at the
\pep2\ collider at SLAC National Laboratory. \pep2\ nominally collides
electrons  and positrons  at  a center-of-mass  (CM) collision  energy
$\sqrt{s}=M_{\Upsilon(4S)}$, producing  pairs of $B$  mesons which are
used to study charge-parity violation.  In the searches presented here
the collision  energy is  tuned to the  $\Upsilon(3S)$ mass,  which is
below the threshold  for $B$ meson pair production.   The width of the
\ups\ is  smaller than that of  the $\Upsilon(4S)$ by  three orders of
magnitude,  and the  branching  fractions for  rare  \ups\ decays  are
therefore larger by $O(10^3)$.   This leads to dramatic enhancement in
the  sensitivity   to  rare  and  exotic   processes,  motivating  the
collection  of  $122\times10^6$ \ups\  decays  at  the  end of  \pep2\
operations.

\section{Search for Lepton-Flavour Violating \ups\ Decays}
In  the SM,  the  rates for  charged  lepton-flavour violating  (CLFV)
processes   are  suppressed   by   $(\Delta(m_{\nu}^2)/M_W^2)^2  \lsim
10^{-48}$~\cite{cite_feinberg,cite_bilenky}    and    are    therefore
unobservable. Here  $\Delta(m_{\nu}^2)$ is the  difference between the
squared  masses of  neutrinos of  different flavour  and $M_W$  is the
charged  weak vector  boson mass.   Several  beyond-the-Standard Model
(BSM) scenarios,  including Supersymmetry and  models with leptoquarks
or compositeness,  predict observable rates for  CLFV processes, which
would therefore provide a clear signal of new physics. These processes
are  generally mediated  by new  particles appearing  in  loops, whose
masses may therefore far exceed the 10.35~GeV CM collision energy. The
previous  constraints  on  CLFV   \ups\  decays  come  from  the  CLEO
experiment,  which  placed  the  95\%  confidence  level  upper  limit
\bfmutau$<2.03\times10^{-5}$~\cite{cite_cleo}.    In   this  analysis,
documented in~\cite{cite_lfv}, upper  limits on \bfetau\ and \bfmutau\
are placed at the $10^{-6}$ level. These results are used to probe new
physics at the TeV mass scale.

The signature  of our signal \etau\  and \mutau\ decays  consists of a
primary high-momentum lepton,  either an electron or muon,  plus a tau
decay in  the opposite hemisphere. The  tau is required to  decay to a
single charged  particle plus  possible additional neutral  pions.  If
the  tau decays  leptonically, we  require that  the tau  daughter and
primary  leptons are  of different  flavour, while  if the  tau decays
hadronically we require one or  two additional neutral pions from this
decay. This leads to four  signal channels, consisting of leptonic and
hadronic tau  decay modes  for the \etau\  and \mutau\  searches.  The
beam-energy-normalized primary lepton  CM momentum $x=p_1/E_B$ is used
to  discriminate between  the  signal and  background processes.   The
signal $x$ distribution is peaked at $x\approx0.97$ since the momentum
of the primary  lepton from the \ltau\ decay is  fixed by the two-body
decay  kinematics.    The  background  is   dominated  by  $\tau$-pair
production,   which  constitutes   an  irreducible   background.   The
$\tau$-pair  $x$  distribution  is   smooth  and  approaches  zero  as
$x\rightarrow  x_{MAX}$, where  $x_{MAX}\approx0.97$ is  the kinematic
endpoint     of     the    lepton     momentum     in    the     decay
$\tau^-\rightarrow\ell^-\bar{\nu}_{\ell}\nu_{\tau}$,  boosted into the
\ups\ rest-frame.  There  is also a contribution to  the \etau\ search
from  Bhabha  events  and   to  the  \mutau\  search  from  $\mu$-pair
events. The  $x$ distributions of  these reducible backgrounds  have a
peaking component near $x=1$.

\begin{figure}[t]
\begin{center}
\includegraphics[width=0.77 \textwidth]{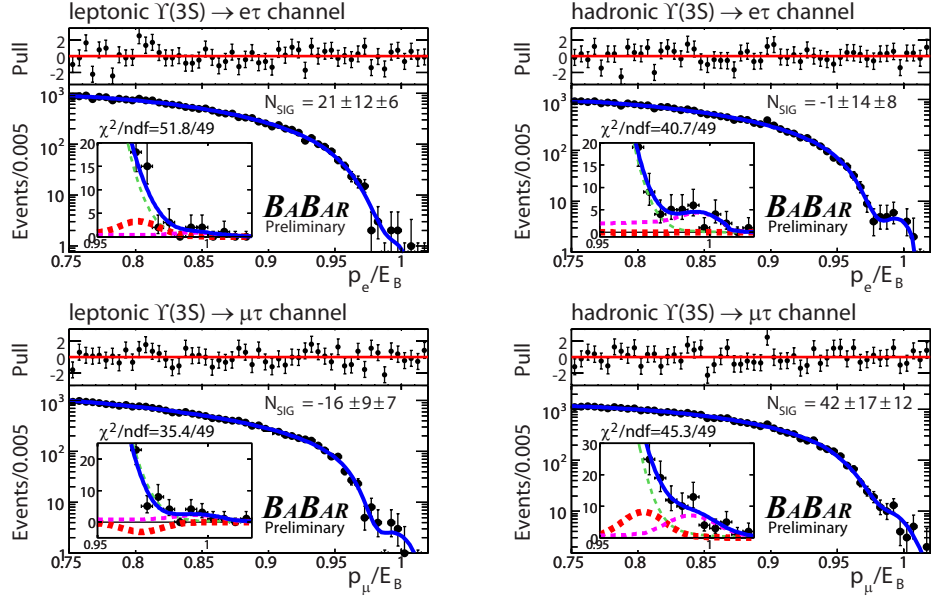}
\end{center}
\vspace{-20pt}
\caption{Fit results  for the  four signal channels,  where `leptonic'
and `hadronic' refer to the  $\tau$ decay mode.  The thin green dashed
line is the $\tau$-pair background PDF, the medium magenta dashed line
is the  Bhabha/$\mu$-pair background  PDF, the thick  red line  is the
signal PDF,  and the solid blue  line is the sum  of these components.
The     inset      shows     a     close-up      of     the     region
$0.95<\mathrm{p_{\ell}/E_B}<1.02$.    The   extracted   signal   yield
$\mathrm{N_{SIG}}$  is displayed with  its statistical  and systematic
uncertainties. }
\label{fits}
\vspace{-10pt}
\end{figure}

All events are required to  pass a set of preselection criteria which
suppress  Bhabha,  $\mu$-pair and  two-photon  processes,  as well  as
beam-gas  interactions.  Events are  then classified  into one  of the
four signal channels based on  the identified particle types and their
CM momenta, and additional kinematic selection criteria are applied to
further  suppress  the   Bhabha  and  $\mu$-pair  backgrounds.   After
selection, an  unbinned, extended maximum likelihood  fit is performed
using the $x$ distributions for the four signal channels individually.
Probability density functions (PDFs) for the signal, Bhabha/$\mu$-pair
and $\tau$-pair  processes are  determined using simulated  events and
$\Upsilon(4S)$ data,  which is not expected to  contain signal events.
A global PDF consisting of the sum of these three components is fitted
to the $x$ distributions, and the yield of each component is extracted
by the fit  as shown in Fig.~\ref{fits}.  The  resulting signal yields
are  all consistent  with zero  within $\pm2.1\sigma$  after including
statistical  and   systematic  uncertainties,  where   the  systematic
uncertainties are  dominated by uncertainties  in the PDF  shapes.  To
extract  the branching  fraction  upper limits,  the  likelihood as  a
function of the branching fractions  is determined for the four signal
channels individually.   The likelihood  functions for the  two \etau\
channels  are  multiplied  to  give  the  combined  \etau\  likelihood
function,  and  likewise  for  the  two \mutau\  channels.   The  90\%
confidence  level  uppers limits  are  determined  by integrating  the
likelihood  functions  $L$ and  finding  $UL$  such that  $\int_0^{UL}
L\,d(BF)$/$\int_0^{\infty} L\,d(BF)=0.9$.   The extracted upper limits
are \bfetau$<5.0\times10^{-6}$, which represents the first upper limit
on this  process, and \bfmutau$<4.1\times10^{-6}$,  which represents a
sensitivity improvement of more than  a factor of four with respect to
the previous upper limit from the CLEO experiment.  These upper limits
can be  used to  constrain new physics  using effective  field theory.
Parameterizing  the   \ltau\  process  ($\ell=e,\mu$)   as  a  generic
$b\bar{b}\ell\tau$   contact   interaction   with  coupling   constant
$\alpha_{\ell\tau}$ and mass  scale $\Lambda_{\ell\tau}$, the branching
fraction     \bfltau\    is     proportional    to     the    quantity
$\alpha_{\ell\tau}^2/\Lambda_{\ell\tau}^4$,  assuming  vector coupling
of            the            $b\bar{b}\ell\tau$            interaction
term~\cite{cite_silagadze,cite_black}.   The upper limits  on \bfltau\
therefore        translate       to       upper        limits       on
$\alpha_{\ell\tau}^2/\Lambda_{\ell\tau}^4$,  which   can  be  used  to
exclude a  region of the  $\Lambda_{\ell\tau}$ vs. $\alpha_{\ell\tau}$
plane.         In        the        strong       coupling        limit
($\alpha_{e\tau}=\alpha_{\mu\tau}=1$), these  results translate to the
90\%  confidence  level  lower  limits  $\Lambda_{e\tau}>1.4$~TeV  and
$\Lambda_{\mu\tau}>1.5$~TeV   on  the  mass   scale  of   new  physics
contributing to CLFV \ups\ decays.

\section{Search for a Low-Mass Scalar in Radiative \ups\ Decays}
We  search for  a  low-mass scalar  particle  $A^0$ in  decays of  the
\ups\ as documented in~\cite{cite_higgs}.  This search is motivated by
theoretical models including  the Next-to-Minimal Supersymmetric Model
(NMSSM), which introduces  a Higgs singlet state in  additional to the
two Higgs  doublets of the  MSSM~\cite{cite_nmssm1,cite_nmssm2}, and a
recent model  which introduces a  light axion~\cite{cite_axion}. These
particles    may    be     produced    in    the    radiative    decay
$\Upsilon(3S)\rightarrow \gamma  A^0$, and the  branching fraction for
this decay is  predicted to be in the  range $10^{-6}-10^{-4}$ for the
given    models.     The   branching    fraction    for   the    decay
$A^0\rightarrow\mu^+\mu^-$  is  predicted to  be  large if  $M_{A^0}<2
m_{\tau}$, leading  to a clean  final state signature consisting  of a
muon  pair and a  photon. This  search is  complementary to  the Higgs
searches performed at the LEP2 and Tevatron experiments.  The searches
performed at  LEP2 are not  sensitive to a  Higgs state whose  mass is
below $2m_b$, since these  experiments searched for the Higgs decaying
to two $b$ jets. The  most stringent constraints on the signal process
come  from   the  CLEO  experiment,  which  placed   the  upper  limit
$BF_{EFF}=BF(\Upsilon(1S)\rightarrow         \gamma         A^0)\times
BF(A^0\rightarrow\mu^+\mu^-)<(1-20)\times10^{-6}$                   for
$M_{A^0}<3.6~\mathrm{GeV/c^2}$~\cite{cite_cleo2}   on   the  effective
signal branching  fraction. Further  motivation for this  search comes
from evidence  of a resonance  structure in the dimuon  invariant mass
distribution of the  decay $\Sigma\rightarrow p\mu^+\mu^-$ observed by
the   HyperCP  experiment~\cite{cite_hypercp}.    Three   events  were
observed   at    $M_{\mu\mu}=214~\mathrm{MeV/c^2}$,   which   may   be
interpreted as a  light scalar decaying to a  muon pair.  Furthermore,
in this  analysis we investigate  the nature of the  $b\bar{b}$ ground
state  $\eta_b$  recently  discovered  at  \babar~\cite{cite_etab}  by
searching for the decay $\eta_b\rightarrow \mu^+\mu^-$.  The branching
for this  decay is not  predicted to be  sizable if the $\eta_b$  is a
$q\bar{q}$ meson state.

\begin{figure}[t]
\begin{center}
\includegraphics[width=0.75 \textwidth]{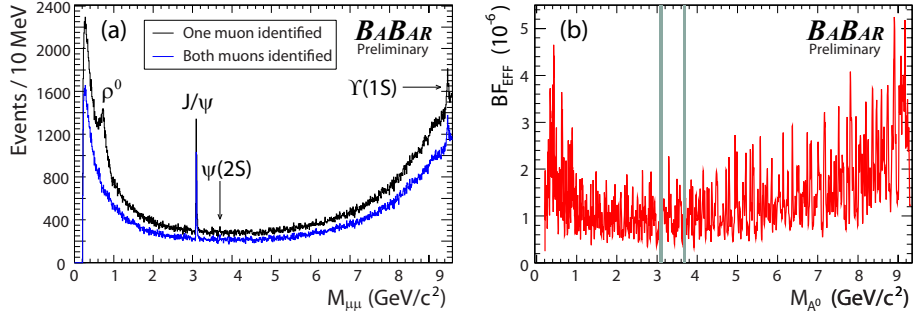}
\end{center}
\vspace{-20pt}
\caption{(a)  Dimuon invariant  mass distribution  for  events passing
selection and (b) 90\% confidence  level upper limits on the effective
branching    fraction    $BF_{EFF}=BF(\Upsilon(3S)\rightarrow   \gamma
A^0)\times  BF(A^0\rightarrow\mu^+\mu^-)$ as a  function of  the $A^0$
mass.}
\label{higgs}
\vspace{-10pt}
\end{figure}

The signature of our events consists of two oppositely-charged tracks,
back-to-back  with a  photon in  the CM  frame. The  photon  energy is
required to satisfy $E_{\gamma}>0.5$~GeV and the total recorded energy
must be  consistent with  $\sqrt{s}$.  The main  background is  due to
$e^+e^-\rightarrow\mu^+\mu^-\gamma$  production,   which  leads  to  a
smooth    dimuon   invariant   mass    distribution   as    shown   in
Fig.~\ref{higgs}a.   The   processes  $e^+e^-\rightarrow  \gamma_{ISR}
\rho^0,~\rho^0\rightarrow \pi^+\pi^-$,  in  which one  of the  charged
pions is misidentified as  a muon, and $e^+e^-\rightarrow \gamma_{ISR}
X,~X\rightarrow \mu^+\mu^-$,  where $X=J/\psi,~\psi(2S),~\Upsilon(1S)$,
contribute additional peaking backgrounds.   The strategy used in this
analysis  is to perform  a series  of maximum  likelihood fits  to the
distribution       of        reduced       mass       defined       by
$m_R=\sqrt{M_{\mu\mu}^2-4m_{\mu}^2}$.   This variable is  used instead
of the dimuon  mass because its distribution is  smoother at low mass.
Using a sliding window of width $\sim300~\mathrm{MeV/c^2}$, $\sim2000$
fits  are performed  in the  range  of reduced  mass corresponding  to
$0.212~\mathrm{GeV/c^2}  \le M_{A^0}  \le  9.3~\mathrm{GeV/c^2}$.  For
each fit,  the signal is modeled  by a peaking  function consisting of
the  sum of two  Crystal Ball  functions~\cite{cite_cb} with  low- and
high-energy tails.   The radiative dimuon  background is modeled  by a
threshold  function  of the  form  $\tanh(\mathrm{poly}(m_R))$ in  the
range  $m_R<230~\mathrm{MeV/c^2}$  and  by  a first-  or  second-order
polynomial for  $m_R>230~\mathrm{MeV/c^2}$.  For the  $\sim2000$ fits,
the   signal    significance   $S$   is    determined   according   to
$S=\mathrm{sign}(N_{SIG})\sqrt{2\log(L_{MAX}/L_0)}$,  where  $N_{SIG}$
is the extracted signal yield, $L_{MAX}$ is the value of the maximized
likelihood function, and $L_0$ is the value of the likelihood function
when the  signal yield is fixed  to zero. The distribution  of $S$ for
the $\sim2000$ fits is found to be well-described by a Gaussian with a
mean  consistent with  zero and  width  consistent with  one, with  no
outliers at  high significance values.  We therefore  conclude that no
statistically significant signal is observed, and the results are used
to    place     the    90\%    confidence     level    upper    limits
$BF_{EFF}=BF(\Upsilon(3S)\rightarrow         \gamma         A^0)\times
BF(A^0\rightarrow\mu^+\mu^-)<(0.25-5.2)\times10^{-6}$   as   shown  in
Fig.~\ref{higgs}b.  These  results represent the best  upper limits to
date   on    the   process   $\Upsilon(3S)\rightarrow    \gamma   A^0,
A^0\rightarrow\mu^+\mu^-$.

We investigate  the HyperCP anomaly by examining  the extracted signal
yield      at     the      reduced      mass     corresponding      to
$M_{\mu\mu}=214~\mathrm{MeV/c^2}$.  The  extracted effective branching
fraction is $BF_{EFF}=(1.2\pm4.3\pm1.7)\times10^{-7}$, where the first
error is statistical and the second is systematic, which is consistent
with zero  and corresponds  to the 90\%  confidence level  upper limit
$BF_{EFF}<8\times10^{-7}$.  We  also search  for dimuon decays  of the
$\eta_b$     using      the     extracted     signal      yield     at
$M_{\mu\mu}=M_{\eta_b}=9.38~\mathrm{GeV/c^2}$.          We        find
$BF(\Upsilon(3S) \rightarrow \gamma \eta_b)\times BF(\eta_b\rightarrow
\mu^+\mu^-) =(0.2\pm3.0\pm0.9)\times10^{-7}$, which is consistent with
zero.  Using  the result from the  \babar\ experiment $BF(\Upsilon(3S)
\rightarrow                                                      \gamma
\eta_b)=(4.8\pm0.5\pm1.2)\times10^{-4}$~\cite{cite_etab}, we determine
$BF(\eta_b\rightarrow\mu^+\mu^-)=(0.0\pm0.6\pm0.2)~\%<0.8~\%$  at 90\%
confidence level.

\section{Conclusions and Outlook}
Decays  of  the  narrow  $\Upsilon$ resonances  provide  an  excellent
laboratory  for  searches for  rare  and  exotic  processes.  We  have
performed a  search for lepton-flavour  violation in \ups\  decays and
placed the  best upper limits to  date, \bfetau$<5.0\times10^{-6}$ and
\bfmutau$<4.1\times10^{-6}$.    These  results   are  used   to  probe
TeV-scale  physics  using  effective   field  theory.   We  have  also
performed a search for a light scalar in radiative decays of the \ups,
and      placed     the     best      upper     limit      to     date
$BF_{EFF}=BF(\Upsilon(3S)\rightarrow         \gamma         A^0)\times
BF(A^0\rightarrow\mu^+\mu^-)<(0.25-5.2)\times10^{-6}$.    We  find  no
evidence to substantiate  the evidence for a light  scalar decaying to
muon pairs  observed by  the HyperCP experiment.   We place  the upper
limit $BF(\eta_b\rightarrow\mu^+\mu^-)<0.8~\%$ on the dimuon branching
fraction  of the  $\eta_b$. All  upper limits  are at  90\% confidence
level.   These results  will be  improved by  extending the  search to
$99\times10^6$  $\Upsilon(2S)$  decays   collected  with  the  \babar\
detector.

\medskip
%{\bf References}

\smallskip

\end{document}